\documentclass[english,pra,showpacs,showkeys,tightenlines,secnumarabic,11pt]{revtex4}
\usepackage[T1]{fontenc}
\usepackage[latin1]{inputenc}
\usepackage{amsmath}
\usepackage{graphicx}
\usepackage{amssymb}
\usepackage{epsfig}
\usepackage{graphics}
\usepackage[mathscr]{euscript}
\usepackage{psfrag}
\usepackage{pstricks}
\usepackage{pst-node}
\usepackage{babel}

\begin{document}
\title{Heavy-quark production in proton-nucleus collisions at
the LHC}
\author{E. Cattaruzza}
\email{ecattar@ts.infn.it}
\author{A. Del Fabbro}
\email{delfabbr@ts.infn.it}
\author{D. Treleani}
\email{daniel@ts.infn.it}
\affiliation{ Dipartimento di Fisica Teorica dell'Universit\`a di Trieste and
INFN, Sezione di Trieste,\\ Strada Costiera 11, Miramare-Grignano,
I-34014 Trieste, Italy.}
\begin{abstract}
A sizable rate of events, with several pairs of heavy-quarks produced contemporarily by
multiple parton interactions, may be expected at very high energies as a consequence of
the large parton luminosities. The production rates are enhanced in hadron-nucleus
reactions, which may represent a convenient tool to study the phenomenon. We compare the
different contributions to $b{\bar b}b{\bar b}$, $c{\bar c}c{\bar c}$ and $b{\bar
b}c{\bar c} $ production due to single and double parton scatterings, in collisions of
protons with nuclei at the CERN-LHC.
\end{abstract}
 \maketitle
\section{Introduction}
The large rates of production of heavy quarks, expected at high energies, may lead to a
sizable number of events, at the CERN-LHC, containing different pairs of heavy quarks,
generated contemporarily by independent partonic collisions. An inclusive cross section
of the order of $10$ $\mu$b may in fact be foreseen for a double parton collision
process, with two $b{\bar b}$ pairs produced in a $pp$ interactions at
$14$TeV\cite{DelFabbro:2002pw}, while the cross section to produce two $c{\bar c}$ pairs
may be of the order of one mb, the contribution of single parton collisions to the
processes being one order of magnitude smaller. All production rates are significantly
enhanced in proton-nucleus collisions, which may offer considerable advantages for
studying multiparton collision\cite{Strikman:2001gz}. Given the large rates expected, the
production of multiple pairs of heavy-quarks should hence represent a convenient process
to study multiparton interactions in $pA$ collisions at the LHC. On the other hand the
mechanism of heavy quarks production is not yet understood satisfactorily also in the
simplest case of nucleon-nucleon collisions, the effects of higher order corrections in
$\alpha_s$ being still a matter of debate. A comprehensive description, of the much more
structured process of heavy quarks production in hadron-nucleus interactions, might hence
be approached after gaining a deeper understanding of the short scale parton-level
dynamics of the process. On the other hand a significant feature of higher order
corrections in $\alpha_s$ is that, for a limited set of physical observables, the whole
effect of higher orders reduces to an approximate rescaling of a lowest order calculation
of heavy quarks production in perturbative QCD \cite{Ryskin:2000bz}. Some of the features
of the process are therefore effectively described by the simplest parton level dynamics
at the lowest order in $\alpha_s$, which lets one speculate that a similar property might
hold also for a much more complex process, as multiple production of heavy quarks.

Taking this optimistic point of view we will attempt, in the present note, to obtain
indications on some properties of $b{\bar b}b{\bar b}$, $c{\bar c}c{\bar c}$ and $b{\bar
b}c{\bar c}$ production in $pA$ interactions at the LHC, by considering the contributions
of the two different interaction mechanisms, the connected $2\to4$ and the disconnected
$(2\to2)^2$ parton processes, which will be effectively described by the lowest order
diagrams in perturbative QCD, while keeping higher order corrections into account by a
simple overall rescaling. Obviously, following this philosophy, all considerations are
necessarily limited to the restricted class of physical observables falling in the
category above.

\section{Production in proton-nucleus collisions}

Quite in general \cite{Paver:1982yp,Braun:2000ua} with the only assumption of
factorization of the hard component of the interaction, the expression of the double
parton scattering cross section to produce two pairs of heavy-quarks is given by

\begin{equation}
\sigma_{(p,A)}^D =\frac{m}{2}\sum_{ij}\int\Gamma_p(x_i,x_j;s_{ij}) \hat{\sigma}(x_i,x_i')
\hat{\sigma}(x_j,x_j')\Gamma_{(N,A)}(x'_j,x'_j;s_{ij})dx_idx_i'dx_jdx_j'd^2s_{ij},
\label{sigmaD}
\end{equation}
where the index $N$ refers to nucleon and $A$ to a nucleus, while the indices $i,j$ to
the different kinds of partons that annihilate to produce a given $q{\bar q}$ pair and
the factor $m/2$ is a consequence of the symmetry of the expression for exchanging $i$
and $j$; specifically $m=1$ for indistinguishable parton processes and $m=2$ for
distinguishable parton processes. The interaction region of a hard process is very small
as compared to the hadron scale, hence in the case of a double parton collision the two
elementary interactions are well localized in transverse space, within the two
overlapping hadrons. The rate of events where two hard collisions take place
contemporarily, in a given inelastic hadron-hadron event, depends therefore on the
typical transverse distance between the partons of the pairs undergoing the multiple
processes. A main reason of interest is hence that double parton scatterings may provide
information on the typical transverse separation between pairs of partons in the hadron
structure. Indeed the non-perturbative input of a double parton collision is the two-body
parton distribution function $\Gamma(x_1,x_2,s_{1,2})$, which depends not only on the
fractional momenta $x_{1,2}$, but also on the relative distance in transverse space
$s_{1,2}$, besides (although not written explicitly to simplify the notation) on the
scales of the two interactions and on the different kinds of partons involved. As a
consequence the double parton scattering cross section is characterized by a linear
dependence on dimensional scale factors, which are related directly with the typical
transverse distances between the various pairs of partons, contributing to the double
scattering process under consideration.

The cross section is simplest when the target is a nucleon and partons are not correlated
in fractional momenta, which may be a sensible approximation in the limit of small $x$.
In such a case the two-body parton distribution may be factorized as
$\Gamma_p(x_i,x_j;s_{ij})=G(x_i) G(x_j) F(s_{ij})$, where $G(x)$ is the usual one-body
parton distribution and $F(s)$ a function normalized to 1 and representing the parton
pair density in transverse space. With this simplifying assumptions the inclusive cross
section to produce two pairs of heavy quarks is written as \cite{Calucci:1999yz}
\begin{equation}
\sigma^D_N=\frac{m}{2}\sum_{ij}\Theta^{ij}\sigma_{i}\,\sigma_{j}\label{sdouble}
\end{equation}
where $\sigma_{i}$ $\sigma_{j}$ represent the inclusive cross sections to produce a
$q{\bar q}$ pair in a hadronic collision, with the indexes $i,\,j$ labelling a definite
parton process. The factors $\Theta^{ij}$ have dimension an inverse cross section and
result from integrating the products of the two-body parton distributions in transverse
space. In this simplified case the factors $\Theta^{ij}$ provide a direct measure of the
different average transverse distances between different pairs of partons in the hadron
structure \cite{Calucci:1999yz, DelFabbro:2000ds}.

The cross section has a more elaborate structure in the case of a nuclear target. The
most suitable conditions to study the phenomenon are those where the nuclear
distributions are additive in the nucleon parton distributions. In such a case one may
express the nuclear parton pair density, $\Gamma_{A}(x'_j,x'_j;s_{ij})$, as the sum of
two well defined contributions, where the two partons are originated by either one or by
two different parent nucleons:

\begin{equation}
\Gamma_{A}(x'_i,x'_j;s_{ij})=\Gamma_{A}(x'_i,x'_j;s_{ij})\Big|_{1}+
\Gamma_{A}(x'_i,x'_j;s_{ij})\Big|_{2}\label{gamma12}\end{equation} and correspondingly
$\sigma_D^A=\sigma_D^A|_1+\sigma_D^A|_2$. The two terms $\Gamma_A|_{1,2}$ are related to
the nuclear nucleon's density. Introducing the transverse parton coordinates
$B\pm\frac{s_{ij}}{2}$, where $B$ is the impact parameter of the hadron-nucleus
collision, one may write

\begin{equation}
\Gamma_{A}(x'_i,x'_j;s_{ij})\Big|_{1,2}=\int
d^2B\gamma_A\Big(x'_i,x'_j;B+\frac{s_{ij}}{2},B-\frac{s_{ij}}{2}\Big)\Big|_{1,2}
\end{equation}
where $\gamma_A|_{1,2}$ are given by
\begin{eqnarray}
&&\gamma_A\Big(x'_i,x'_j;B+\frac{s_{ij}}{2},B-\frac{s_{ij}}{2}\Big)\Big|_{1}=
\Gamma_N(x'_i,x'_j;s_{ij})T(B)\cr
&&\gamma_A\Big(x'_i,x'_j;B+\frac{s_{ij}}{2},B-\frac{s_{ij}}{2}\Big)\Big|_{2}=G_N(x'_i)G_N(x'_j)
T\Big(B+\frac{s_{ij}}{2}\Big)T\Big(B-\frac{s_{ij}}{2}\Big) \label{gammaa}
\end{eqnarray}
with $T(B)$ is the nuclear thickness function, normalized to the atomic mass number $A$
and $G_N$ nuclear parton distributions divided by the atomic mass number.

In the simplest additive case, the first term in Eq.(\ref{gamma12}) obviously gives only
a rescaling of the double parton distribution of a isolated nucleon
\begin{equation}
\Gamma_{A}(x'_i,x'_j;s_{ij})\Big|_{1}=\Gamma_N(x'_i,x'_j;s_{ij})\int d^2BT(B)
\end{equation} and the resulting contribution to the double parton scattering cross
section is the same as in a nucleon-nucleon interaction, apart from an enhancement factor
due to the nuclear flux, which is given by the value of the atomic mass number $A$:
\begin{equation}
\sigma_A^D\Big|_1=A\sigma_N^D. \label{sigma1}
\end{equation}
The $\sigma_A^D|_2$ term has more structure. In this case the integration on the relative
transverse distance between the partons of the interacting pairs, $s_{ij}$, involves both
the projectile and two different target nucleons:

\begin{equation}
\int ds_{ij}\Gamma_p(x_i,x_j;s_{ij})T\Big(B+
\frac{s_{ij}}{2}\Big)T\Big(B-\frac{s_{ij}}{2}\Big). \label{gt}
\end{equation}
As one may notice the expression (\ref{gt}) depends on two very different scales, the
hadron radius $r_p$ and the nuclear radius $R_A$. A usual approximation in $pA$
interactions is to consider the limit $r_p\ll R_A$, where one may use the approximation
\begin{equation}
T\Big(B\pm\frac{s_{ij}}{2}\Big)\simeq T(B), \end{equation} which allows one to decouple
the integrations on $s_{ij}$ and on $B$. One hence obtains:

\begin{equation}
\sigma_A^D\Big|_2=\frac{1}{2}\sum_{ij}\int G_p(x_i,x_j)\hat{\sigma}(x_i,x_i')
\hat{\sigma}(x_j,x_j')G_N(x_i')G_N(x_j')dx_idx_i'dx_jdx_j'\int d^2BT^2(B), \label{sigma2}
\end{equation}
where
\begin{equation}
G_p(x_i,x_j)=\int d^2s_{ij}\Gamma_p(x_i,x_j;s_{ij}).\end{equation}

Remarkably the two terms $\sigma_A^D\Big|_1$ and $\sigma_A^D\Big|_2$ have very different
properties. In fact, while the correct dimensionality of $\sigma_A^D\Big|_1$ is provided
by transverse scale factors related to the {\it nucleon} scale, eqs.(\ref{sdouble},
\ref{sigma1}), the analogous dimensional factor is provided in $\sigma_A^D\Big|_2$ by the
{\it nuclear} thickness function, which is at the second power in Eq.(\ref{sigma2}),
being two the target nucleons involved in the interaction.

As pointed out in ref. \cite{Strikman:2001gz}, while on general grounds
$\sigma^D_{(p,A)}$ depends both on the longitudinal and transverse parton correlations,
the $\sigma_A^D\Big|_2$ term depends solely on the longitudinal momentum fractions
$x_i,x_j$ so that, when the $\sigma_A^D\Big|_2$ term is isolated, one has the capability
of measuring the longitudinal and, a fortiori, also the transverse parton correlations of
the hadron structure in a model independent way.

Although the two contributions may be defined in a more general case, the separation of
the cross section in the two terms $\sigma_A^D\Big|_1$ and $\sigma_A^D\Big|_2$ is most
useful in the regime of additivity of the nuclear structure functions, which may not be a
bad approximation for a sizable part of the kinematical regime of heavy-quarks production
at the LHC. In the case of a central calorimeter with the acceptance of the ALICE
detector ($|\eta|<0.9$), the average value of momentum fraction of the initial state
partons, in a $pp\to b{\bar b}b{\bar b}$ process, is $\langle x\rangle\approx
6\times10^{-3}$ while for $pp\to c{\bar c}c{\bar c}$ one finds $\langle x\rangle\approx
2\times10^{-3}$ . With a cut of $5$ GeV in the transverse momenta of the $b, \,c$ quarks
one obtains $\langle x\rangle\approx 10^{-2}$ and $\langle x\rangle\approx
8\times10^{-3}$ respectively, while a cut of $20$ GeV gives $\langle x\rangle\approx
3\times10^{-2}$. In the case of $2.5<\eta<4$ and without any cut in $p_t$, the average
values of momentum fraction are $\langle x\rangle\approx 3\times10^{-2}$ and $\langle
x\rangle\approx 9\times10^{-3}$ \cite{Charm-bottom:2003}. Deviations from additivity at
low $x$ are less than $10\%$ for $x\ge2\times10^{-2}$ \cite{Amaudruz:1995tq} and,
although increasing with the atomic mass number, non additive corrections should hence be
at most a $20\%$ effect, in above the kinematical regimes.

As mentioned in the introduction, heavy quarks production is characterized by a non
trivial dynamics, in such a way that also next-to-leading corrections to the lowest order
term in perturbative QCD are not sufficient for an exhaustive description of the
inclusive spectra, which most likely need an infinite resummation to be evaluated. After
comparing the results of different approaches to heavy quarks production (as NLO QCD and
$k_t$-factorization) one nevertheless finds that, in a few cases, the whole effect of
higher orders is to a large extent just an approximate rescaling of the results obtained
by a lowest order evaluation in perturbative QCD. When limiting the discussion to an
accordingly restricted set of physical observables, the whole effect of higher order
corrections to heavy quark production is hence summarized by a single number, the value
of the so-called $K$-factor defined as:
\begin{equation}
K=\frac{\sigma(q\bar q)\,}{\sigma_{LO}(q\bar q)}
\end{equation}
where $\sigma(q\bar q)$ is the inclusive cross section for $q\bar q$ production and
$\sigma_{LO}(q\bar q)$ the result of the lowest order calculation in pQCD. By evaluating,
with the $k_t$-factorization approach, rapidity and pseudorapidity distributions of
$b{\bar b}$ and of $c{\bar c}$ production, in $pp$ collisions at the center-of-mass
energy of ${\sqrt s}=5.5$ TeV, within $|\eta|<0.9$ and $2.5<\eta<4$, with different
choices of heavy-quark masses and of the factorization and renormalization scales, one
finds a result not incompatible with a lowest order calculation in perturbative QCD
rescaled by the factors $K$ shown in Tables
[\ref{K-factor1},\ref{K-factor2},\ref{K-factor3}].

To obtain the values of the K-factor in Table [\ref{K-factor1}],
in the evaluation of the cross section with the
$k_t$-factorization approach, we have set the factorization scale
equal to the invariant mass of the parton process $\hat s$ and the
renormalization scale equal to the virtuality of the initial state
gluons, while, for the cross section at the lowest order in pQCD,
we used as a scale factor the transverse mass of the produced
quarks. In Tables [\ref{K-factor2}, \ref{K-factor3}] we rather
used in both cross sections the average of the squared transverse
masses of the produced quarks and the heavy-quark mass as scale
factors\cite{Charm-bottom:2003}.

With a fixed choice of factorization and renormalization scales, one obtains a
substantial increase of the value of the $K$ factor when decreasing the value of the
heavy quark mass, the increase being larger for $c \bar c$ than for $b \bar b$
production. With our different choices for the scale factors we obtain variations of the
$K$ factor within $4\%$ for $b \bar b$ and $15\%$ for $c \bar c$ production.

To discuss $q{\bar q}q{\bar q}$ production in $pA$ collisions, while remaining in a
kinematical regime where non additive corrections to the nuclear structure functions are
not a major effect, we have limited all considerations to physical observables, where
higher orders may be taken into account by a simple rescaling of the lowest order
calculation. For $ q{\bar q}q{\bar q}$ production, where only results of three level
calculations are up to now available, we have further assumed that the value of the
$K$-factor in the $2\to4$ parton process is the same as in the $2\to2$ process
\cite{DelFabbro:2002pw, DelFabbro-Double:2003}.
 We have hence evaluated the various contributions to the cross section in the case of a
 central,
$|\eta|<0.9$, and of a forward calorimeter, $2.5<\eta<4$, where the approximate
expression of the cross section in Eq.(\ref{sdouble}) may not be an unreasonable
approximation. As the process is dominated by gluon fusion, the expression of the cross
section in Eq.(\ref{sdouble}) may be limited to a single term only. For the corresponding
dimensional scale factor we used the value reported by the CDF measurement of double
parton collisions \cite{Abe:1997bp,Abe:1997xk} and to evaluate the cross section at the
lowest order in pQCD we used the MRS99 parton distributions \cite{mrs99}, with
factorization and renormalization scale equal to the transverse mass of the produced
quarks. The cross sections of the $2\to4$ processes have been evaluated generating the
matrix elements of the partonic amplitudes with MadGraph~\cite{madgraph} and HELAS~
\cite{helas}. For the mass of the bottom quark we used the central values $m_b=4.6$ and
$m_c=1.4$ GeV. The multi-dimensional integrations have been performed by
VEGAS~\cite{vegas} and the lowest order pQCD cross sections have been finally multiplied
by the $K$ factors of Table [\ref{K-factor1}].
\section{Results}
A major result of the present analysis is that the effects induced by the presence of the
nucleonic degrees of freedom, in double parton scatterings with a nuclear target, cannot
be reduced to the simple shadowing corrections of the nuclear parton structure functions,
whose effect is to decrease the cross section as a function of $A$. In the case of double
parton collisions, the main effect of the nuclear structure is represented by the
presence of the $\sigma_A^D|_2$ term in the cross section, which scales with a different
power of $A$ as compared to the single scattering contribution, producing an additive
correction to the cross section.

To emphasize the resulting ``anomalous'' dependence of the double parton scattering cross
section, as a function of the atomic mass number, we have plotted in fig.[\ref{ratio}]
the ratio $\sigma_2^D/(\sigma_1^D+\sigma^S)$, as a function of $A$. The ratio represents
the contribution to the cross section of the processes where two different nuclear target
nucleons are involved in the interaction, scaled with the contribution where only a
single target nucleon is involved. The dependence on the atomic mass number of the latter
terms of the cross section, namely the single ($\sigma^S$) and the double ($\sigma_1^D$)
parton scattering terms against a single nucleon in the nucleus, is the same of all hard
processes usually considered, where nuclear effects may be wholly absorbed in the
shadowing corrections to the nuclear structure functions. The contribution to the cross
section of the $\sigma_2^D$ term is, on the contrary, ``anomalous'', involving two
different target nucleons in the interaction process. The ratio above hence represents
the relative weights of the ``anomalous'' to the ``usual'' contributions to the double
parton scattering cross
section on a nuclear target. \\
The plots in Fig.[\ref{eta-forward}] show the rapidity distribution in a forward
calorimeter ($2.5\le\eta\le4$) of a heavy quark produced in an event with $b{\bar b}
b{\bar b}$, $c{\bar c} c{\bar c}$ and $b{\bar b} c{\bar c}$ respectively: one-nucleon
(dashed lines) and two-nucleon contributions (continuous lines) in the case of a heavy,
{\it Au}, (higher curves) and of a light nucleus, {\it O}, (lower curves). In each case
considered continuous and dashed curves differ essentially only by a rescaling, showing
that the effect of the single parton scattering term (the $2\to4$ parton process) is
negligible in this kinematical regime. The different contributions to the cross section
for $b\bar b b\bar b$, $c\bar c c\bar c$, $b\bar b c\bar c$ production, due to
interactions with a single (dashed line) or with two different target nucleons (dot
line), are shown in
Fig.[{\ref{bottom.pB.s5.5},\ref{charm.pB.s5.5},\ref{bottom.charm.pB.s5.5}] as a function
of the atomic mass number in the case of a central and of a forward calorimeter, with two
different choices of cuts ($p_t=0,10\,GeV$) on the outgoing quark-transverse momenta. The
continuous line is the sum of the two contributions. As one may see, by introducing a
cutoff in $p_t$ one enhances the single scattering contribution and the relative role of
$\sigma_2^D$ is decreased.\\
To have some indication on the overall uncertainty of our
estimates we have plotted in Fig.[\ref{factorization-mass}] the
range of values obtained for the integrated cross sections, for
$b\bar b b\bar b$, $c\bar c c\bar c$ and $b\bar b c\bar c$
production, within $|\eta|\le.9$, as a function of the atomic mass
number, when making the different choices described above. In each
panel we report the different choices corresponding to the extreme
values of the cross section. The indication obtained in this way
is that the overall cross
section is roughly determined within a factor three.\\
Summarizing the large size of the cross section of heavy-quark
production in hadron-nucleus collisions at the LHC (one may expect
values of the order of $5-10\,mb$ for charm production) suggests
that the production of multiple pairs of heavy quarks is fairly
typical at high energies, hence representing a convenient channel
to study multiple parton interactions. A rather direct feature,
which is a simplest prediction and then a test of the interaction
mechanism described above, is the ``anomalous'' dependence on $A$.
The effects induced by the presence of the nucleonic degrees of
freedom in the nuclear structure are in fact not limited, in this
case, to the shadowing corrections to the nuclear structure
functions usually considered, which cause a limited {\it decrease}
(not larger than $20\%$, in the kinematical regime considered
here) of the cross section for a hard interaction in
hadron-nucleus collisions. When considering double parton
scatterings, all nuclear effects can be exhausted in the shadowing
corrections to the nuclear structure functions in the
$\sigma_A^D|_1$ term only. The dominant effect of the nuclear
structure is on the contrary due to the presence of the
$\sigma_A^D|_2$ term in the cross section, which scales with a
different power of $A$ as compared to single scattering term,
giving rise to a sizably larger correction, with opposite sign as
compared to the shadowing correction, namely to an {\it increase}
of the cross section, which may become larger than $100\%$ for a
heavy nucleus.
\section{Acknowledgment}This work was
partially supported by the Italian Ministry of University and of Scientific and
Technological Researches (MIUR) by the Grant COFIN2003.

\newpage
\begin{table}[t]
\begin{tabular}{|c|c|c|c|}
\hline
\multicolumn{4}{|c|}{$\boldsymbol{\mu_F^2=\hat{s}\qquad\mu_R^2=q^2}\,\,$\bf{(gluon virtuality)}}\\
\hline \hline
$\,\,\boldsymbol{m_b\,(GeV)}\,\,$&$\,\,\boldsymbol{K_b=\sigma(b\bar{b})/\sigma_{LO}(b\bar{b})}\,\,$&
$\,\,\boldsymbol{m_c\,(GeV)}\,\,$&$\,\,\boldsymbol{K_c=\sigma(c\bar{c})/\sigma_{LO}(c\bar{c})}\,\,$\\
\hline
$4.25$&$\,\,\,5.9\,\,\,$&$1.2$&$\,\,\,7.0\,\,\,$\\
\hline
$4.5$&$\,\,\,5.7\,\,\,$&$1.4$&$\,\,\,6.6\,\,\,$\\
\hline
$4.75$&$\,\,\,5.5\,\,\,$&$1.6$&$\,\,\,6.1\,\,\,$\\
\hline
\end{tabular}
\caption{$K$ factor values for different choices of the mass of
the heavy quarks. In the $k_t$ factorization approach the cross
section has been evaluated by using $\hat s$ as factorization
scale and the gluon virtuality as renormalization scale; when
evaluating the cross section at the lowest order in pQCD the
scales have been set equal to the transverse mass of the heavy
quarks\cite{Charm-bottom:2003}} \label{K-factor1}
\end{table}
\begin{table}
\begin{tabular}{|c|c|c|c|}
\hline
\multicolumn{4}{|c|}{$\boldsymbol{\mu_F^2=\mu_R^2=\mu_0^2=(m_{t,Q}^2+m_{t,\bar{Q}}^2)/2}$}\\
\hline \hline
$\,\,\boldsymbol{m_b\,(GeV)}\,\,$&$\,\,\boldsymbol{K_b=\sigma(b\bar{b})/\sigma_{LO}(b\bar{b})}\,\,$&
$\,\,\boldsymbol{m_c\,(GeV)}\,\,$&$\,\,\boldsymbol{K_c=\sigma(c\bar{c})/\sigma_{LO}(c\bar{c})}\,\,$\\
\hline
$4.25$&$\,\,\,6.1\,\,\,$&$1.2$&$\,\,\,6.2\,\,\,$\\
\hline
$4.5$&$\,\,\,5.9\,\,\,$&$1.4$&$\,\,\,5.9\,\,\,$\\
\hline
$4.75$&$\,\,\,5.8\,\,\,$&$1.6$&$\,\,\,5.6\,\,\,$\\
\hline
\end{tabular}
\caption{$K$ factor values for different choices of the masses of the heavy quarks. In
the calculation of the cross sections, factorization and renormalization scales have been
set equal to the average of the squared transverse masses of the heavy
quarks\cite{Charm-bottom:2003}} \label{K-factor2}
\end{table}
\begin{table}
\begin{tabular}{|c|c|c|c|}
\hline
\multicolumn{4}{|c|}{$\boldsymbol{\mu_F^2=\mu_R^2=m_Q^2}$}\\
\hline \hline
$\,\,\boldsymbol{m_b\,(GeV)}\,\,$&$\,\,\boldsymbol{K_b=\sigma(b\bar{b})/\sigma_{LO}(b\bar{b})}\,\,$&
$\,\,\boldsymbol{m_c\,(GeV)}\,\,$&$\,\,\boldsymbol{K_c=\sigma(c\bar{c})/\sigma_{LO}(c\bar{c})}\,\,$\\
\hline
$4.25$&$\,\,\,5.7\,\,\,$&$1.2$&$\,\,\,8.1\,\,\,$\\
\hline
$4.5$&$\,\,\,5.6\,\,\,$&$1.4$&$\,\,\,7.9\,\,\,$\\
\hline
$4.75$&$\,\,\,5.4\,\,\,$&$1.6$&$\,\,\,7.3\,\,\,$\\
\hline
\end{tabular}
\caption{$K$ factor values for different choices of the masses of the heavy quarks. In
the calculation of the cross sections, factorization and renormalization scales have been
set equal to the mass of the heavy quark. \cite{Charm-bottom:2003}} \label{K-factor3}
\end{table}
\begin{figure}[t]
\begin{center}
\includegraphics[scale=0.7,angle=270]{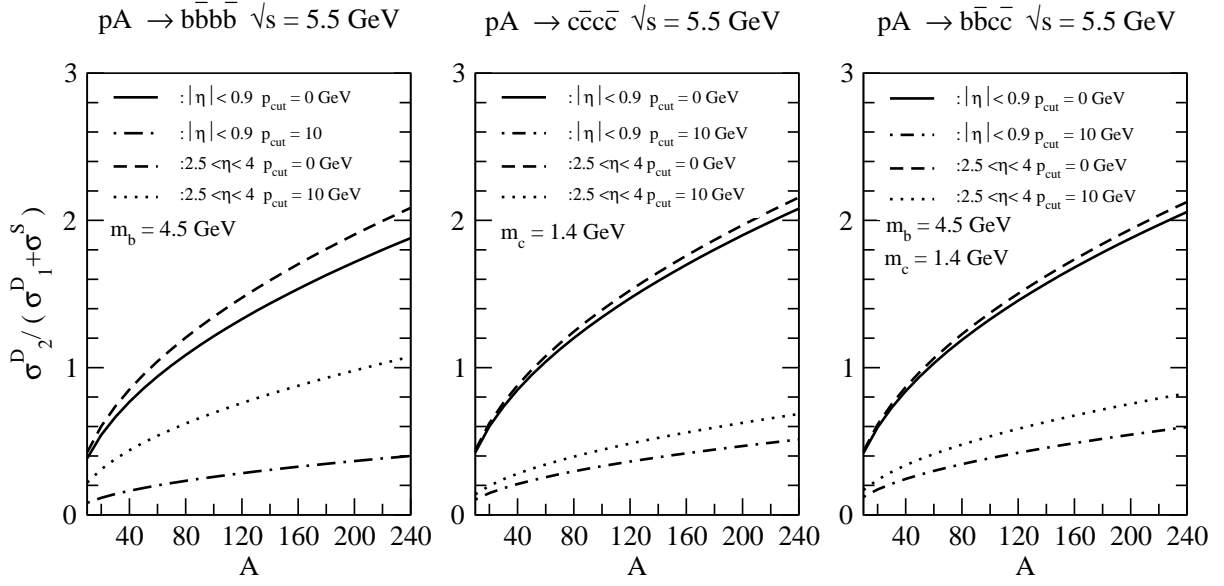}
\caption {Relative weights of the terms with ``anomalous'' and ``usual'' $A$-dependence
in the double scattering cross section for $b\bar b b \bar b,\,c\bar c c \bar c,\,b\bar b
c \bar c $ production.} \label{ratio}
\end{center}
\end{figure}
\begin{figure}[b]
\begin{center}
\includegraphics[scale=0.6,angle=270]{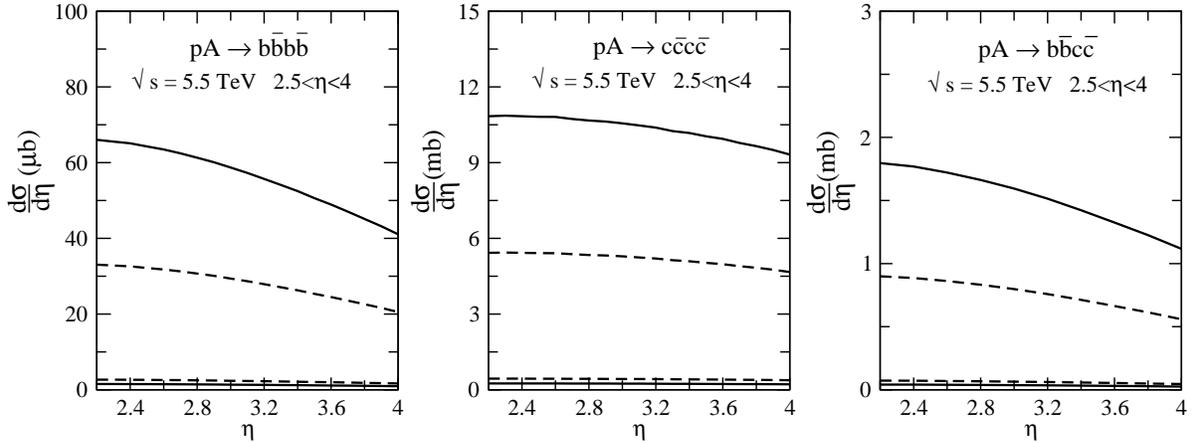}
\caption{Pseudorapidity distribution in a forward calorimeter of a heavy quark produced
in events with $b\bar b b \bar b$, $c\bar c c \bar c$ and $b\bar b c \bar c$: one-nucleon
(dashed lines) and two-nucleon contributions (continuous lines) in the case of a heavy
(higher curves) and of a light nucleus (lower curves).} \label{eta-forward}
\end{center}
\end{figure}
\begin{figure}[t]
\begin{center}
\includegraphics[scale=0.5,angle=270]{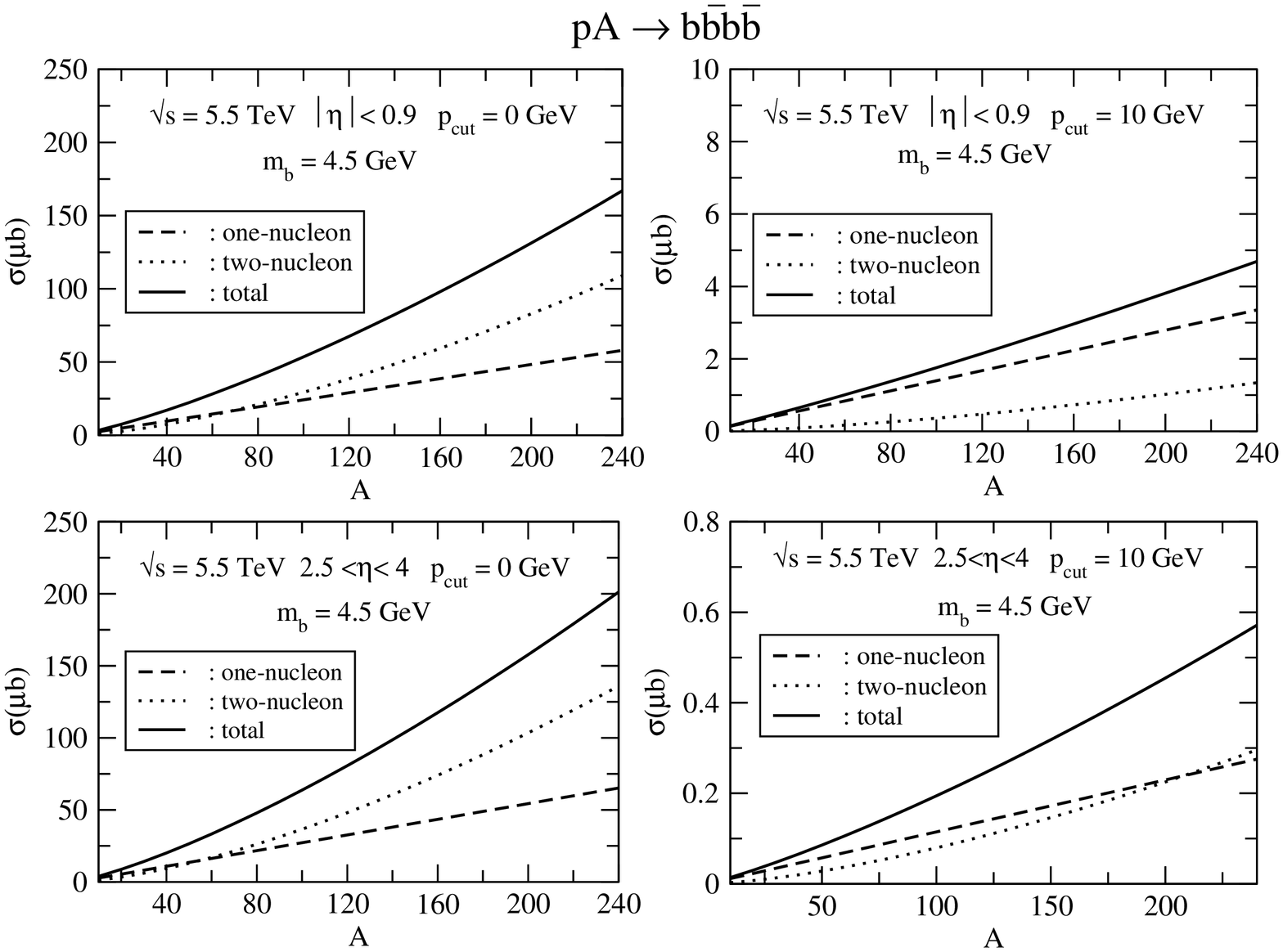}
\caption{ Different contributions to the cross section of $b\bar b b \bar b$ production
in a central and in a forward calorimeter as a function of $A$. Cross sections without
any cut in $p_t$ (left figures) and after applying a cut of $10$ GeV in the transverse
momenta of each produced heavy-quark (right figure): one-nucleon (dashed lines), two
nucleons contribution (dotted lines) and total cross section (continuous lines).}
\label{bottom.pB.s5.5}
\includegraphics[scale=0.5,angle=270]{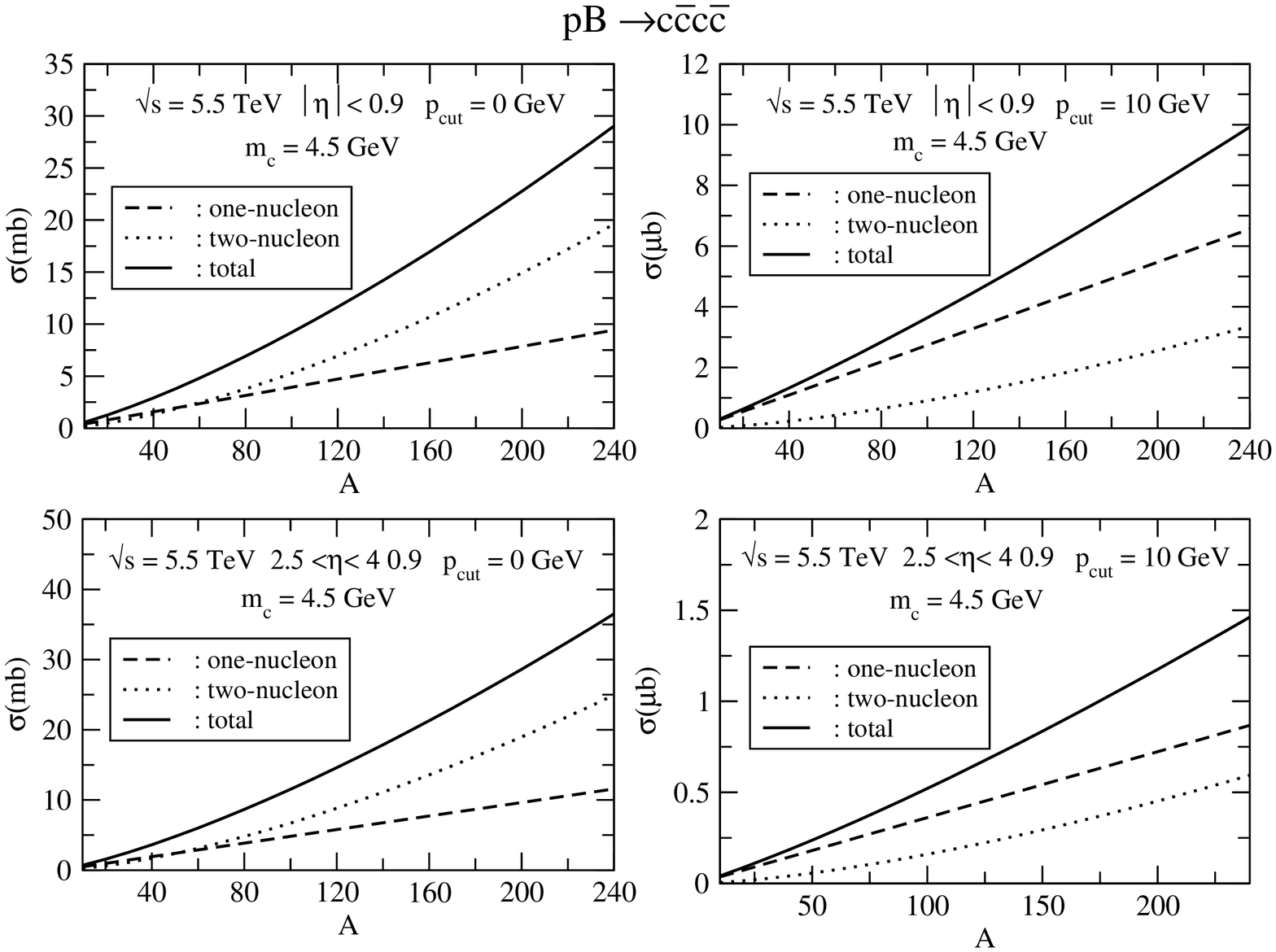}
\caption{Different contributions to the cross section of $c\bar c c \bar c$ production in
a central and in a forward calorimeter as a function of $A$. Cross sections without any
cut in $p_t$ (left figures) and after applying a cut of $10$ GeV in the transverse
momenta of each produced heavy-quark (right figure): one-nucleon (dashed lines), two
nucleons contribution (dotted lines) and total cross section (continuous lines).}
\label{charm.pB.s5.5}
\end{center}
\end{figure}
\begin{figure}
\begin{center}
\includegraphics[scale=0.5,angle=270]{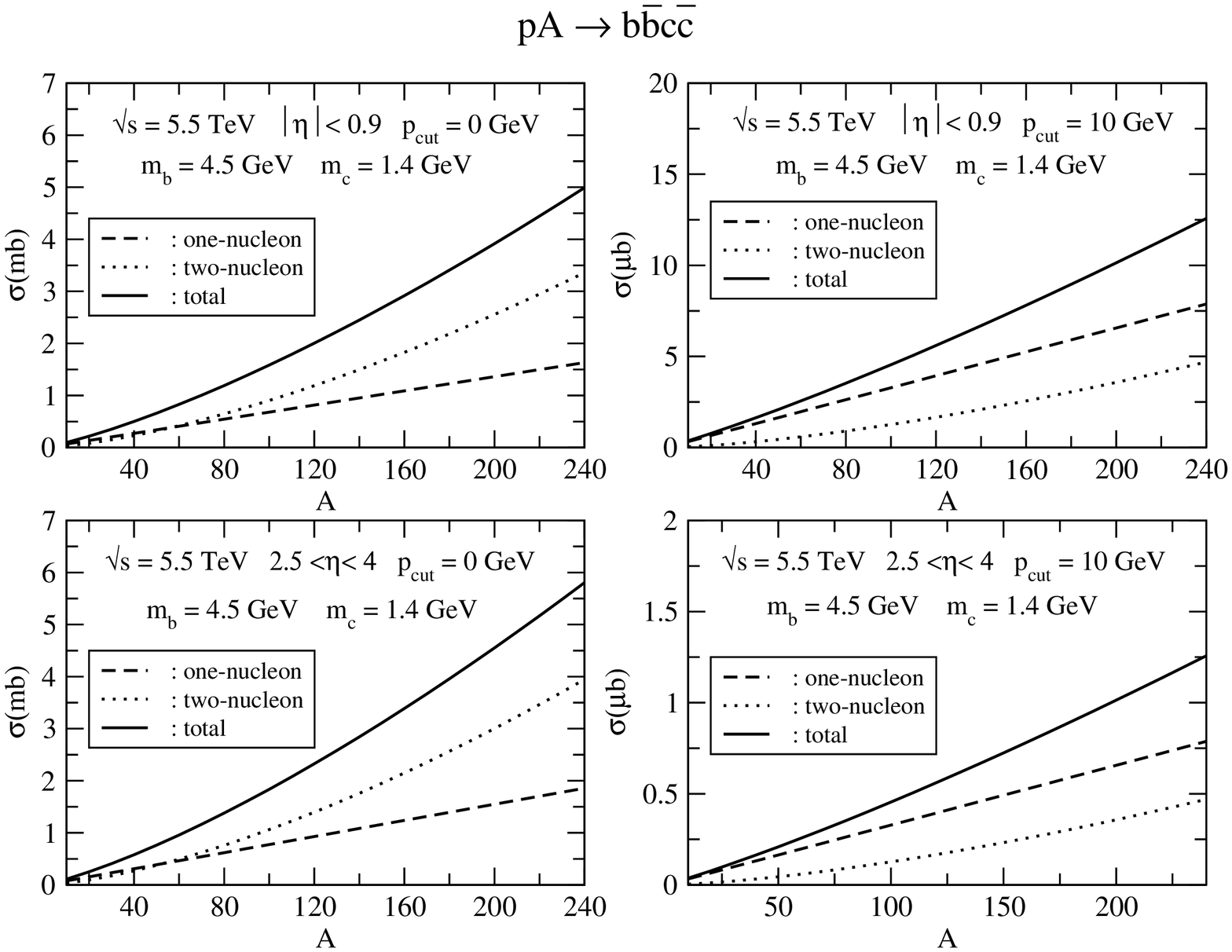}
\caption{ Different contributions to the cross section of $b\bar b c \bar c$ production
in a central and in a forward calorimeter as a function of $A$. Cross sections without
any cut in $p_t$ (left figures) and after applying a cut of $10$ GeV in the transverse
momenta of each produced heavy-quark (right figure): one-nucleon (dashed lines), two
nucleons contribution (dotted lines) and total cross section (continuous lines).}
\label{bottom.charm.pB.s5.5}
\end{center}
\end{figure}
\begin{figure}
\begin{center}
\includegraphics[scale=0.6,angle=270]{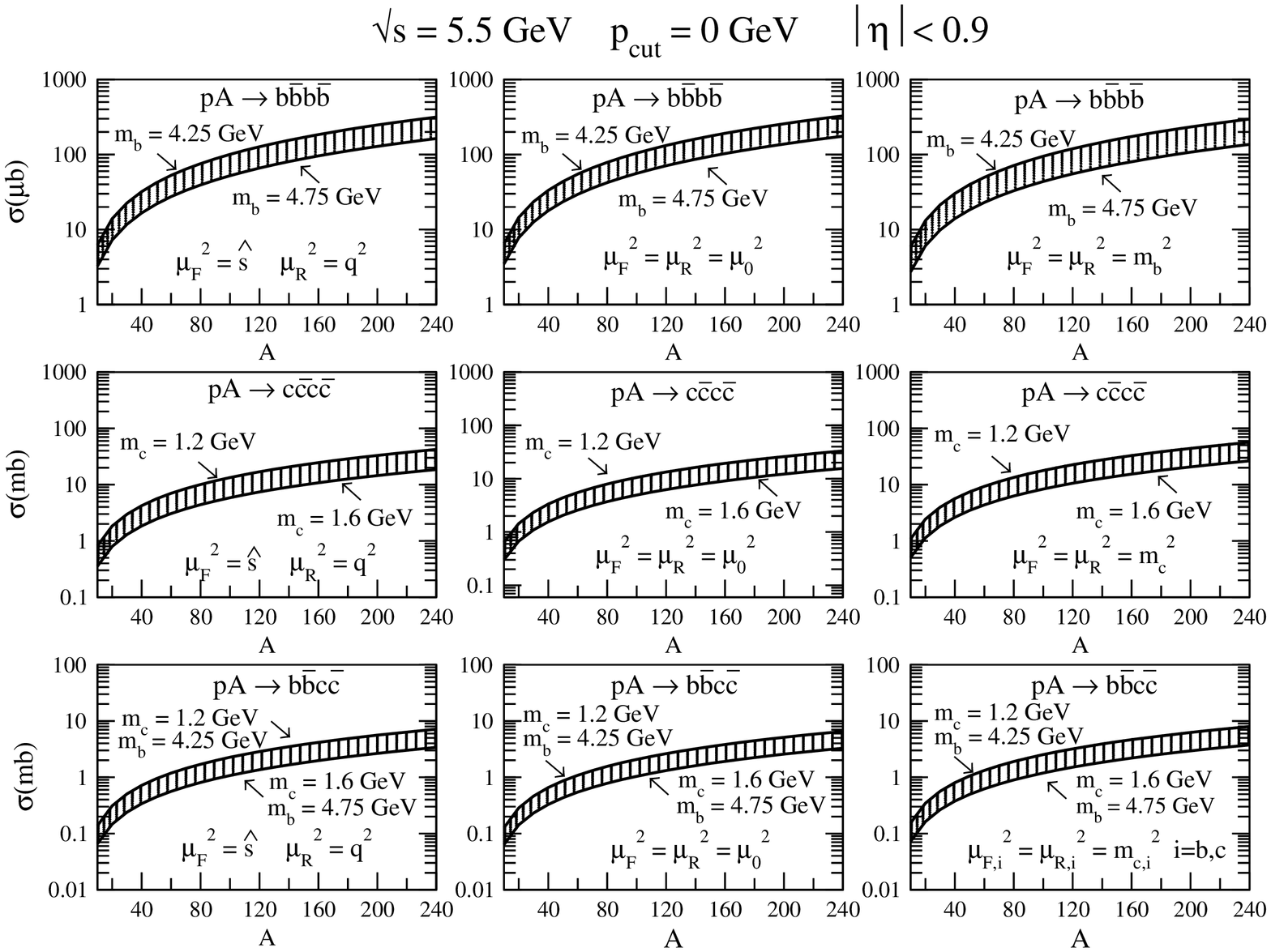}
\caption{Production cross section of $b\bar b b\bar b,\,c\bar c c\bar c $ and $b\bar b
c\bar c$ in a central calorimeter for different choices of quark-masses and factorization
scales. Lower and upper curves, correspond respectively to the lower and upper values of
the heavy quark masses considered in our calculation. } \label{factorization-mass}
\end{center}
\end{figure}

\begin{thebibliography}{99}
\bibitem{DelFabbro:2002pw}
A.~Del Fabbro and D.~Treleani,
Phys. Rev. {\bf D66} (2002) 074012 [arXiv:hep-ph/0207311].
\bibitem{Strikman:2001gz}
M.~Strikman and D.~Treleani, Phys.\ Rev.\ Lett.\  {\bf 88}, 031801 (2002)
[arXiv:hep-ph/0111468].
\bibitem{Ryskin:2000bz}
M.~G.~Ryskin, A.~G.~Shuvaev and Y.~M.~Shabelski,
Phys.\ Atom.\ Nucl.\  {\bf 64} (2001) 1995 [Yad.\ Fiz.\  {\bf 64} (2001) 2080]
[arXiv:hep-ph/0007238].

\bibitem{Paver:1982yp}
N.~Paver and D.~Treleani,
Nuovo Cim.\ A {\bf 70}, 215 (1982).
\bibitem{Braun:2000ua}
M.~Braun and D.~Treleani,
Eur.\ Phys.\ J.\ C {\bf 18}, 511 (2001) [arXiv:hep-ph/0005078].

\bibitem{Calucci:1999yz}
G.~Calucci and D.~Treleani,
Phys.\ Rev.\ {\bf D60} (1999) 054023.

\bibitem{DelFabbro:2000ds}
A.~Del Fabbro and D.~Treleani, Phys.\ Rev.\ D {\bf 63}, 057901 (2001)
[arXiv:hep-ph/0005273].
\bibitem{Charm-bottom:2003}
N.~Carrer and A.~Dainese, [arXiv:hep-ph/0311225], M.~Bedjidian et {\it al}
[arXiv:hep-ph/0311048], J.~Baines et {\it al} [arXiv:hep-ph/0003142].
\bibitem{DelFabbro-Double:2003}
A.~Del Fabbro and D.~Treleani, [arXiv:hep-ph/0301178].
\bibitem{Amaudruz:1995tq}
P.~Amaudruz {\it et al.}  [New Muon Collaboration],
Nucl.\ Phys.\ B {\bf 441}, 3 (1995) [arXiv:hep-ph/9503291].
\bibitem{Abe:1997bp}
F.~Abe {\it et al.}  [CDF Collaboration], Phys.\ Rev.\ Lett.\ {\bf 79}, 584 (1997).
\bibitem{Abe:1997xk}
F.~Abe {\it et al.}  [CDF Collaboration],
Phys.\ Rev.\ D {\bf 56}, 3811 (1997).
\bibitem{mrs99}
A.D.Martin, R.G.Roberts, W.J.Stirling and R.S.Thorne, Eur.\ Phys.\ J. {\bf C14} (2000)
133.
\bibitem{madgraph}
T.Stelzer and W.F.Long, Comp.Phys.Comm. {\bf 81}, 357 (1994);
\bibitem{helas}
E.Murayama, I.Watanabe and K.Hagiwara, HELAS: HELicity Amplitude Subroutines for Feynman
Diagram Evaluations, KEK report 91-11, January 1992;
\bibitem{vegas}
G.~P.~Lepage,
J.\ Comput.\ Phys.\ {\bf 27}, 192 (1978).
\end{thebibliography}
\end{document}